\documentclass[sigconf]{acmart}
\usepackage[utf8]{inputenc}
\usepackage[T1]{fontenc}
\usepackage[draft,inline,nomargin,index]{fixme}
\usepackage{amssymb,amsmath,graphicx,comment,multirow,microtype}
\usepackage{booktabs,color,xspace,balance,csquotes}
\usepackage{mathtools}
\newcommand\defeq{\stackrel{\mathclap{\normalfont\mbox{def}}}{=}}

\fxsetup{theme=colorsig,mode=multiuser,inlineface=\itshape,envface=\itshape}
\FXRegisterAuthor{cw}{acw}{Christo}
\FXRegisterAuthor{ps}{aps}{Piotr}
\FXRegisterAuthor{wz}{awz}{Wesley}
\FXRegisterAuthor{am}{aam}{Alan}

\newcommand{\para}[1]{{\vspace{7pt} \bf \noindent #1 \hspace{10pt}}}
\newcommand{\topic}{\vspace{7pt}\noindent}

\newenvironment{packed_enumerate}{
\begin{enumerate}
  \setlength{\itemsep}{2pt}
  \setlength{\parskip}{0pt}
  \setlength{\parsep}{0pt}
  \setlength{\topsep}{2pt}
  \setlength{\itemindent}{0pt}
}{\end{enumerate}}

\newcommand{\eg}{e.g.,\ }
\newcommand{\etal}{et~al.\xspace}
\newcommand{\ie}{i.e.,\ }

\newcommand{\aka}{a.k.a.\ }

\newcommand{\fulltitle}{Quantifying the Impact of User Attention\\on Fair Group Representation in Ranked Lists}
\newcommand{\fullkeywords}{information retrieval; group fairness; ranked lists}

\definecolor{linkColor}{RGB}{6,125,233}
\definecolor{left}{RGB}{100,149,237}
\definecolor{right}{RGB}{205,92,92}
\hypersetup{%
  pdftitle={\fulltitle},
  pdfauthor={},
  pdfkeywords={\fullkeywords},
  bookmarksnumbered,
  pdfstartview={FitH},
  colorlinks,
  citecolor=black,
  filecolor=black,
  linkcolor=black,
  urlcolor=linkColor,
  breaklinks=true,
  hypertexnames=false
}

\copyrightyear{2019}
\acmYear{2019} 
\setcopyright{iw3c2w3}
\acmConference[WWW '19 Companion]{Companion Proceedings of the 2019 World Wide Web Conference}{May 13--17, 2019}{San Francisco, CA, USA}
\acmBooktitle{Companion Proceedings of the 2019 World Wide Web Conference (WWW '19 Companion), May 13--17, 2019, San Francisco, CA, USA}
\acmPrice{}
\acmDOI{10.1145/3308560.3317595}
\acmISBN{978-1-4503-6675-5/19/05}

\renewcommand\footnotetextcopyrightpermission[1]{}

\begin{document}

\renewcommand{\sectionautorefname}{\S}
\renewcommand{\subsectionautorefname}{\S}
\renewcommand{\subsubsectionautorefname}{\S}

\title{\fulltitle}

\author{Piotr Sapiezynski}
\affiliation{%
  \institution{Northeastern University}
  %\city{Boston}
  %\state{MA}
}
\email{p.sapiezynski@northeastern.edu}

\author{Wesley Zeng}
\affiliation{%
  \institution{Northeastern University}
  %\city{Boston}
  %\state{MA}
}
\email{zeng.wes@husky.neu.edu}

\author{Ronald E. Robertson}
\affiliation{%
  \institution{Northeastern University}
  %\city{Boston}
  %\state{MA}
}
\email{robertson.ron@husky.neu.edu}

\author{Alan Mislove}
\affiliation{%
  \institution{Northeastern University}
  %\city{Boston}
  %\state{MA}
}
\email{amislove@ccs.neu.edu}

\author{Christo Wilson}
\affiliation{%
  \institution{Northeastern University}
  %\city{Boston}
  %\state{MA}
}
\email{cbw@ccs.neu.edu}

\begin{abstract}
In this work, we introduce a novel metric for auditing group fairness in ranked lists.
Our approach offers two benefits compared to the state of the art.
First, we offer a blueprint for modeling of user attention. 
Rather than assuming a logarithmic loss in importance as a function of the rank, we can account for varying user behaviors through parametrization.
For example, we expect a user to see more items during a viewing of a social media feed than when they inspect the results list of a single web search query.
Second, we allow non-binary protected attributes to enable investigating inherently continuous attributes (\eg political alignment on the liberal to conservative spectrum) as well as to facilitate measurements across aggregated sets of search results, rather than separately for each result list.
By combining these two elements into our metric, we are able to better address the human factors inherent in this problem. We measure the whole sociotechnical system, consisting of a ranking algorithm and individuals using it, instead of exclusively focusing on the ranking algorithm.
Finally, we use our metric to perform three simulated fairness audits.
We show that determining fairness of a ranked output necessitates knowledge (or a model) of the end-users of the particular service. 
Depending on their attention distribution function, a fixed ranking of results can appear biased both in favor and against a protected group.
\end{abstract}




\maketitle
%\pagenumbering{arabic}
\section{Introduction}
\label{sec:intro}

The exponential growth of information available online has necessitated the development of Information Retrieval (IR) algorithms that decide what content is \textit{relevant} to users. For example, upwards of 5.5 billion searches are conducted on Google every day~\cite{google2016}, and in response to each, Google filters and sorts a list of $\sim$15 results~\cite{robertson2018auditing}. Similarly, over 1.4 billion people visit Facebook daily~\cite{facebook-2018q2} and scroll through a list of content from friends and advertisers deemed relevant by the News Feed algorithm. Finally, tens of millions of worker profiles are available on LinkedIn, filtered and sorted when recruiters search for prospective employees.

%Although many IR systems are benign, 
Recently, a concern has been growing that even seemingly benign IR systems may negatively impact people. 
It has been shown that algorithms can reflect societal biases~\cite{barocas2016big}, and ranking mechanisms are no different. Kay~\etal found that Google Image Search returned images portraying men and women in stereotypical roles in response to occupation-related queries, 
%(\eg men are doctors, women are nurses), 
and that these results reinforced
%peoples' perceptions of 
stereotypical gender roles~\cite{kay-2015-URG}. 
Others have examined partisan slant in search results~\cite{kulshrestha-2017-cscw,diakopoulos-2018-vote,robertson2018auditing} in light of user studies demonstrating that partisan search results can significantly influence voting behavior~\cite{epstein2015search,epstein-cscw-2017}. Lastly, two studies have 
%critically 
examined the relationships between gender, race, and ranking of job seekers on employment websites~\cite{hannak2017bias,chen2018investigating}, where systematic biases that push members of protected classes into lower ranks could result in the loss of employment opportunities and earnings~\cite{kim-2017-datadisc}. 

Only recently have researchers started addressing the problem of fairness in ranked outputs. From the IR side, this includes novel ranking algorithms that aim to achieve \textit{representational parity} (\aka \textit{group fairness}): the ranker is required to assign a certain fraction of top ranks to people in the protected or minority class)~\cite{zehlike2017fa,celis-2017-arxiv,singh2018fairness,biega2018equity}. 
%These algorithms offer vendors the opportunity to develop IR systems that are explicitly designed to eliminate \textit{disparate treatment} and \textit{disparate impact}~\cite{pedreshi-2008-discrimination}, for example. 
From the \textit{algorithm auditing} side (\ie investigators who look for fairness problems in black-box systems)~\cite{sandvig2014auditing}, Yang and Stoyanovich introduced metrics for quantifying whether the outputs from a given search engine are group-fair~\cite{yang2017measuring}. This enables auditors to 
%critically 
examine real-world search engines and hopefully hold them accountable for producing unfair outputs.

However, there is a reoccurring challenge in the extant literature on fair ranking: how to model user attention? Eye-tracking studies and click-stream data 
%consistently 
show that users do not distribute their attention evenly over ranked lists of information~\cite{granka-2004-eye,guan-2007-eye,murphy-2006-clicking,dimitrov-2017-wikipedia}. %Further, click-stream data reveals that the placement of items on a ranked list strongly influences the probability of users clicking on them~\cite{murphy-2006-clicking,dimitrov-2017-wikipedia}.
This unequal distribution of attention must be taken into account when designing fair IR systems and evaluating whether a given IR system is fair. The trouble is that the distribution of attention for a given search engine may be unknown, since it varies based on the user interface of the service (\eg pagination boundaries) and context (\eg searching for a specific movie trailer versus searching for a new employee). 

Most of the previous work on fairness in ranked lists has assumed logarithmic discounting of attention~\cite{yang2017measuring,singh2018fairness}. 
%For example, Yang and Stoyanovich assume that importance drops logarithmically with the results' page number (thus each element on a given page has the same importance)~\cite{yang2017measuring}. Singh and Joachims use the standard exposure drop-off with ranking of of $1/\log(1 + j)$, where $j$ is the rank~\cite{singh2018fairness}.
However, because of its flattening shape for low ranks, logarithmic drop-off is impractical for modeling attention. For example, modeling attention this way would mean assuming that on a list of 100 results, the sum of attention given to last eight results is bigger than the attention paid to the first. Biega~\etal use a geometric distribution instead, but do not investigate the consequences of varying its steepness~\cite{biega2018equity}.

In this work, we extend the literature on fair ranking by introducing a novel metric for measuring group fairness in ranked outputs. Our metric, the \textit{Viable-$\Lambda$ Test}, is designed for auditors and answers two questions: (1) does there exist a distribution of user attention $P(\Lambda)$ such that the output of a search engine is group fair, and (2) if so, what is the parameterization $\Lambda = (\lambda_1, ..., \lambda_m)$ of this distribution? In contrast to prior work that attempts to ``score'' the fairness of a ranking algorithm~\cite{yang2017measuring}, our metric fundamentally re-frames the question of fairness to involve the consumer of the ranking and their attention. If the fitted model of attention $P(\Lambda)$ does not match empirical observations of user attention in the given search engine, then the system does not achieve representational parity.

Overall, our paper makes three key contributions: 
\begin{packed_enumerate}
    \item We introduce a novel metric, the Viable-$\Lambda$ Test that binds the usage patterns of a list to the measurement of fairness. 
    \item We enable fairness measurements in situations with class assignment uncertainty, results aggregation, multiple protected classes, and continuous protected variables. 
    \item We evaluate the Viable-$\Lambda$ Test on data from three real-world services: a resume search engine, a dating service, and a web search engine. Our results demonstrate that the choice and parameterization of the attention function can lead to dramatically different conclusions about whether (and how) the rankings are biased.
\end{packed_enumerate}

Note that it is not possible to determine with certainty whether a given set of search results are biased without knowing the true attention distribution function for users of the corresponding service.
Therefore, our work should {\em not} be seen as an audit study, 
but rather a showcasing of a metric that can be used by the operators or internal auditors of these services to ensure fair delivery of results.

The remainder of the paper is organized as follows.
In Section~\ref{sec:background} we introduce related work on fairness in ranked lists and auditing of ranking algorithms.
In Section~\ref{sec:methods} we provide a detailed description of the mechanics of the proposed metric.
In Section~\ref{sec:design}we explain how the parameters of our metric should be set and interpreted based on context of its use.
In Section~\ref{sec:case_studies} we analyze three case studies: a hiring service, a dating service, and Google search.
In Section~\ref{sec:limitations} we explain the limitations of our approach.
Section~\ref{sec:discussion} suggests the directions for further research, and Section~\ref{sec:conclusion} concludes the article.
\section{Background}
\label{sec:background}

\para{In Pursuit of Fairness.} As use of large, observational datasets has proliferated, so have concerns that systems leveraging this data may have a negative impact on people. The machine learning community has mapped the legal concepts of \textit{disparate treatment} and \textit{disparate impact}  to \textit{direct} and \textit{indirect discrimination} by algorithms, respectively~\cite{pedreshi-2008-discrimination,calders-2009-BCI,hajian-2013-methodology}. 
%Direct discrimination occurs when an algorithm explicitly uses protected class attributes as features, while indirect occurs when there are proxies in the data that correlate with protected class attributes~\cite{calders-2009-BCI,hajian-2013-methodology}. 
Zafar~\etal introduced the concept of \textit{disparate mistreatment} to refer to situations where false positives and negatives are not equally distributed across subpopulations~\cite{zafar-2015-fairness,zafar-2017-FBD}. Such situations have been shown to occur \eg in pre-trial assessments~\cite{angwin2016machine} and academic performance predictors~\cite{sapiezynski2017academic}.

While direct discrimination can be corrected by removing the protected attributes from the data, indirect discrimination is more challenging to address. Dwork~\etal proposed two potential objectives for mitigating indirect discrimination: under \textit{individual fairness}, similar people should be treated similarly by the algorithm, while under \textit{group fairness} subpopulations should be treated equivalently to the whole population~\cite{dwork2012fairness}. There is a large and growing literature on how to achieve these objectives in machine learning-based classifiers~\cite{kamiran-2010-DAD,kamishima-2011-fairness,luong-2011-KIS,kamishima-2012-FCP,dwork2012fairness,kamiran-2009-discrimination,calders-2009-BCI,kamiran-2010-prefsample,calders-2010-three,zliobaite-2011-icdm,kamiran-2012-decision,calders-2013-controlling,hajian-2013-methodology,feldman-2015-CRD}. 

\para{Fairness in Ranking.} Achieving fairness in IR systems has received less attention in the academic community. One challenge is that research on fair classification does not necessarily generalize to the ranking context.
%because of data dependencies in the output of the system. For example, adjusting the classification of an item or items does not necessarily impact the classification of all other items; conversely, adjusting the rank of an item or items in search results \textbf{does} impact the rank of all adjacent items. 
A second challenge is accounting for \textit{order effects}~\cite{murdock-1962-serial}, \ie the well-established tendency of human beings to pay more attention to items at the top\footnote{We use ``top'' and ``high'' to refer to the numerically lowest ranks in lists, \eg rank one, in keeping with the norms of the IR literature~\cite{craswell-wdsm-08,joachims-wdsm-17}.} of a ranked list.

A few methods for generating group-fair search results have been proposed. Zehlike~\etal leverage randomization by positing that a given ranked output is fair if it could have been generated by a random Bernoulli process~\cite{zehlike2017fa}. Celis~\etal propose a more general approach allowing the user to specify the fairness constraints~\cite{celis-2017-arxiv}. Unfortunately, neither of these take user attention into account: their methods do not distinguish between different orderings of a set as long as a minimum fraction of items from the minority class are presented at each rank. In contrast, Singh and Joachims argue that even if the ranking itself is unbiased, small differences in placement may lead to large discrepancies of attention~\cite{singh2018fairness}. % because of nonlinear distribution of exposure
 %They propose a probabilistic approach for generating group-fair rankings in which the items from majority and minority groups receive, on average, similar amount of exposure. 
Biega~\etal point out that any single ranking of similarly relevant items is individually unfair because of the uneven distribution of attention~\cite{biega2018equity}. Therefore, they propose achieving individual equity (attention corresponding to relevance) within a certain number of realizations by systematic reshuffling of the list.

\para{Algorithm-in-the loop Approach.} Most of the work we discussed so far focus on measuring or correcting the algorithm without explicitly involving its users. In contrast, Green and Chen emphasize the need for considering the whole sociotechnical system~\cite{green2019disparate}; they show that rather than focusing solely on the bias in an automated risk-assessment system, one needs to include contextual information on the system is actually used by judges and how it affects their decisions.

\para{Auditing Search Engines.} There is a growing body of work from the algorithm auditing~\cite{sandvig2014auditing} community that aims to measure whether real-world search engines are fair and unbiased. Kay~\etal found that Google Image Search presented results that were stereotypically gendered~\cite{kay-2015-URG}, while Hannak~\etal and Chen~\etal showed that search engines on employment websites were not group fair with respect to race and gender~\cite{hannak2017bias,chen2018investigating}. Audit studies have also examined the political partisanship of search results from Twitter and Google~\cite{kulshrestha-2017-cscw,diakopoulos-2018-vote,robertson2018auditing}.

An open challenge for the auditing community is selecting appropriate metrics for assessing whether search engine results are group fair. For example, Kay~\etal only looked at simple metrics like average representation that fail to take order effects into account~\cite{kay-2015-URG}. Other audits have used group representation in top $K$ ranks~\cite{hargreaves2019fairness}, logarithmic discounting~\cite{hannak2017bias,chen2018investigating,yang2017measuring} and linear normalization by rank~\cite{kulshrestha-2017-cscw,robertson2018auditing} to model the decay of attention. In this work, we argue that these ad hoc methods do not accurately model users' attention, and may lead to incorrect conclusions about (un)fairness of IR systems.

% Yang and Stoyanovich designed a trio of metrics to measure fairness in ranked lists~~\cite{yang2017measuring}. Each implements the following general procedure: for $k$ pages of search engine output with $n$ results on each, calculate the difference between the expected and observed number of minority class items on each page, and logarithmically discount the pages' importance from one to $k$. 
% %This metric assumes that the overall fraction of minority items in the population are known.
% Unfortunately, these metrics do not generalize well to search engines that do not paginate their output, and the authors provide little guidance on how to choose the threshold that numerically delineates fair from unfair scores. Moreover, while these metrics account for decreasing user attention across pages, they do not model attention within each page.
% Finally, similarly to other mentioned approaches, these metrics only allow for binary group assignment without covering cases where belonging to a group is probabilistic (\eg when there are multiple realizations of the ranking) or inherently continuous rather than dichotomous (\eg partisan bias).

\section{Methods}
\label{sec:methods}

In this section, we introduce a novel metric for measuring group fairness in search results. This metric, the Viable-$\Lambda$ test, combines existing research from algorithmic auditing, IR, and human-computer interaction to address the human factors inherent in this problem.

\subsection{Overview}

Suppose we are given an ordered list of search results $R = [ r_1, \dots, r_n ]^T$. %We assume that users browse the search results beginning with item $r_1$ and then proceed down to $r_k$ where $k \le n$. 
Our goal is to measure the representation of some target property $p$ that is shared by each $r_i \in R$. The metric proposed in this paper requires the auditor to specify the following five elements:
(1)~an \textit{alignment vector} $L_R$, (2)~an \textit{attentional weight vector} $W_R$, (3)~a \textit{population estimator} $\hat{p}$, (4)~a \textit{distance metric} $d$, and (5)~a \textit{maximum allowable distance} $\delta_{\textrm{max}}$.

Below we briefly introduce these elements and explain their role in the Viable-$\Lambda$ Test.
%(\cwerror{Missing fig --> see \autoref{fig:search_engine_results} for an illustrative example of each}). 
In Section~\ref{sec:design} we explain the design choices behind each element and how to allocate them appropriately in practice.

\topic The \textbf{alignment vector} $L_R$ is a vector of probability distributions $[l_1,  \dots, l_n]^T$ that describe the group membership (or \textit{alignment}) of $r_i$ with respect to the target property $p$. The subscript $_R$ indicates that $L_R$ has the same length as $R$, and that each $l_i \in L_R$ corresponds to the alignment of each respective $r_i \in R$. $l_i$ can be either discrete or continuous.

\topic The \textbf{attentional weight vector} $W_R$ is a probability vector $[ \omega(r_1), \dots, \omega(r_n)]^T$ that models the relative user attention allocated to each $r_i \in R$. While it is difficult to determine the exact distribution of $W_R$, we can make assumptions about its \textit{shape}. Formally, $W_R \sim P(\Lambda)$ where $P$ is a family of $n$-truncated discrete probability distributions with an unknown set of true parameters $\Lambda$.

We calculate the expected cumulative exposure $E_R$ of group representation in $R$ by taking the dot product of $L_R$ and $W_R$:
 \begin{equation}
    E_R=L_R^T \cdot W_R
    \label{eq:exposure}
\end{equation}
Note that $E_R$ is a probability distribution with the same domain as each $l_i \in L_R$ (\ie a distribution describing the target property $p$).

\topic The \textbf{population estimator} $\hat{p}$ is a probability distribution that estimates the true demographics of $p$. For reasons described in~\autoref{sec:estimating-p-hat}, the following formula is often a well-motivated choice:
%for $\hat{p}$:
\begin{equation}
    \hat{p} \defeq \bar{L_R} = \sum_{r_i \in R} \frac{l(r_i)}{n}
    \label{eq:phat}
\end{equation}
where $\hat{p}$ has the same domain as $l_i$.  

\topic The \textbf{distance metric} $d$ quantifies the statistical difference between the probability distributions $E_R$ and $\hat{p}$. % We motivate possible choices for $d$ in \autoref{sec:d-and-delta-min}.

\topic The \textbf{maximum acceptable distance} $\delta_{\textrm{max}}$ is the threshold that separates group fair from unfair search results. It is chosen in a context-dependent manner in conjunction with $d$. As we describe in \autoref{sec:d-and-delta-min}, $d$ is essentially a statistical significance test, and $\delta_{min}$ is the test statistic threshold. Both components draw inspiration from the principles of traditional sampling statistics.

\subsection{The Viable-\texorpdfstring{$\Lambda$}{Lambda} Test for Representational Parity}

Given $L_R$, $\hat{p}$, $d$, and $\delta_{\textrm{max}}$, we define group exposure as $E_R=L_R^T \cdot W_R$. Assume that $W_R \sim P(\lambda_1, \dots, \lambda_m)$ where $P$ is an $n$-binned discrete probability distribution with an unknown set of true parameters $\Lambda = (\lambda_1, ..., \lambda_m)$ in some domain space $D$ with known bounds. Then $R$ is unfair if:
\begin{packed_enumerate}
\item $\nexists \lambda \in D$ such that $d(e_r,\hat{p}) < \delta_{\textrm{max}}$, \ie there is no way to parameterize the attention distribution such that representational parity is attained; and
\item For $\Lambda$ satisfying the above condition, $W(\Lambda)$ matches reasonable expectations and data about true user behavior. 
%\item $\exists \Lambda \in D$ such that $d(E_R,\hat{p}) < \delta_{\textrm{max}}$, but $P(\Lambda)$ does not match the true distribution of attention exhibited by users of the given search engine.
\end{packed_enumerate}
% \begin{figure*}
%     \centering
%     \includegraphics[width=1\textwidth]{figures/seng.pdf}
%     \caption{Calculating fairness of a set of a search engine result lists, where the protected attribute is represented by a symbol (circle, square, or a triangle).}
%     \label{fig:search_engine_results}
% \end{figure*}

\section{Design Choices}
\label{sec:design}

This section describes the key components of Viable-$\Lambda$ and the motivating decisions behind their design. 
%[in addition to some HCI and psychological rootings behind the choices and the advantages that they offer in terms of representing group fairness. ]

%%%%%%%%%%%%%%%%%%%%%%%%%%%%%%%%%
%%%%%       Alignment       %%%%%
%%%%%%%%%%%%%%%%%%%%%%%%%%%%%%%%%

\subsection{The Alignment Vector \texorpdfstring{$L_R$}{LR}}
\label{sec:L-R}

% Let $L_R$ denote $[l(r_1), \dots, l(r_n)]^T$, \ie a \textit{vector of probability distributions} describing each $r_i$'s alignment respectively. 

\para{Alignment as a Probability Distribution.} Given an ordered list of result items $R = [ r_1, \dots, r_n ]^T$, we use an \textit{alignment function} $\iota$ to map each result $r_i$ to a probability distribution describing its \textit{alignment} in terms of the target property $p$ (\eg race, gender, political alignment, etc.).
As a motivating example, consider a resume search engine~\cite{chen2018investigating} that returns a list of job candidates $R$ in response to a query (\eg ``software engineer''). Suppose we want to model the gender alignment $p$ of these results; namely, $p$ is a discrete probability distribution across three classes: `Male,' `Female,' and `Unknown.' If the search results \textit{explicitly} display the gender of each candidate $r_i$, then we can define alignment as follows:
\begin{equation}
    l_i \defeq 
    \begin{cases}
        \text{\{Female: 1.0, Male: 0.0, Unknown: 0.0\}} & \text{if $r_i$ is female} \\
        \text{\{Female: 0.0, Male: 1.0, Unknown: 0.0\}} & \text{if $r_i$ is male} \\
        \text{\{Female: 0.0, Male: 0.0, Unknown: 1.0\}} & \text{otherwise}
    \end{cases}
\end{equation}

In realistic scenarios, defining $l_i$ may not be so trivial. Continuing our example, resume search engines typically do not explicitly state the gender or race of job candidates. However, recruiters (and auditors) may still be able to infer them using other information, like a user's profile picture, given name, etc. 
Most existing approaches to measuring fairness assume that all $l_i$ are explicitly known and binary (canonically between protected and non-protected classes $S$ and $S^C$)~\cite{dwork2012fairness,yang2017measuring}.
However, this assumption has not generalized to empirical studies.
By using a probability distribution rather than a binary indicator, we cover the cases where the class assignment is ambiguous, where there are multiple classes, or there are multiple realizations of the ranking.

%%%%%%%%%%%%%%%%%%%%%%%%%%%%%%%%%
%%%%%       Attention       %%%%%
%%%%%%%%%%%%%%%%%%%%%%%%%%%%%%%%%

\subsection{Attentional Weights \texorpdfstring{$W_R$}{WR}}
\label{sec:W-R}

$W_R$ encapsulates the well-documented fact that search engine users do not treat all search results equally~\cite{granka-2004-eye,guan-2007-eye,richardson-www-07,craswell-wdsm-08}. For example, the first result on Google Search is estimated to receive approximately 30\% of all clicks, and the results on the first page account for approximately 90\% of clicks~\cite{keane2008people}. This observation remains true to a large extent even if the order of the search results is inverted~\cite{keane2008people}.

\para{Modelling Attentional Weights $\omega$.} $W_R$ is a probability vector that models the amount of user attention that each $r_i \in R$ receives. $W_R  = [ w_1, \dots, w_n ]^T$, where $w_i = \omega(r_i) \in [0, 1]$ is the output of a weighting function $\omega$ applied to $r_i$.
%Although $\omega$ and $\iota$ are applied to the same search results, they measure two distinct properties of $r_i$, and are thus influenced by different data fields. $\iota$ is influenced by predictors of $p$, while 
$\omega$ is influenced by (1) the user interface design of the search engine (for example pagination or highlighted results) and (2) the use context of the search engine. For an intuition on the latter, consider that a user may only view several web search results before finding an acceptable answer to their query, whereas that same user might view dozens of resumes from a resume search engine if they are tasked with hiring a new employee. 

\para{Fitting $W_R$ from Empirical Data.} A tempting way to approximate $W_R$ is to use empirical data such as organic Click Volume (CV) and Click-Through Rate (CTR), widely used in digital marketing and by search engine proprietors.
Unfortunately, there are three serious impediments to using an empirically derived $W_R$. First, click data is often proprietary, and thus unavailable to an external auditor. Second, click data is only an approximation for what users see in search results. Truly measuring user attention might require expensive eye tracking studies~\cite{granka-2004-eye,guan-2007-eye,richardson-www-07,craswell-wdsm-08}. Third, as we noted above, the way users distribute their attention over search results depends on website design and use context. Thus, although eye-tracking studies are available for sites like Google Search~\cite{granka-2004-eye,guan-2007-eye,richardson-www-07,craswell-wdsm-08}, the data may not be applicable to other search engines.

\para{Potential Choices for $W_R$.} In algorithmic auditing, we need a way to measure user attention without (1) having access to the vendor's analytic data, and (2) having to conduct multiple eye tracking studies. We partially address this issue by making assumptions about the shape of $W_R$. Formally, we assume that $W_R \sim P(\Lambda)$ for some discrete truncated probability distribution $P$. $W_R$ should meet two criteria:
\begin{packed_enumerate}
    \item $W_R$ is an $n$-truncated discrete probability distribution.
    \item Higher-ranked results receive substantially more attention than lower-ranked ones; \ie for reasonably large $n$, $\omega(r_1) \gg\ \omega(r_i)$ as $i \xrightarrow{}n$.
\end{packed_enumerate}
% Some families of distributions that fit the above criteria are presented in~\autoref{fig:distributions} and include:
% \begin{packed_itemize}
%     \item The \textbf{Truncated Geometric Distribution}. This probability distribution assumes a constant rate of decay in $W_R$ dictated by a success parameter $p$. Under a geometric interpretation of attention, the user reads $R$ beginning with $r_1$ and continues serially at a fixed rate $p$, ending their search (\ie by finding a satisfying result or quitting) at some $r_i$.
%     \item The \textbf{Truncated Log-series Distribution}. A potential issue with the geometric distribution is that it assumes that the success parameter $p$ is independent of $i$. This does not account for decay in search item \textit{relevancy} as $i$ increases. Log-series allows us to capture this intuition by introducing a secondary decay factor based on the rank $i$.
%     \item The \textbf{Truncated Discrete Pareto distribution}~\cite{kozubowski2015discrete}. Zipfian and Pareto relationships are found in many scientific fields including linguistics, economics, and data compression. Although a Pareto model of attention may have interesting implications about how the human mind interacts with search results, we will leave this as an area for future research.
% \end{packed_itemize}

\begin{figure}
    \centering
    \includegraphics[width=1\linewidth]{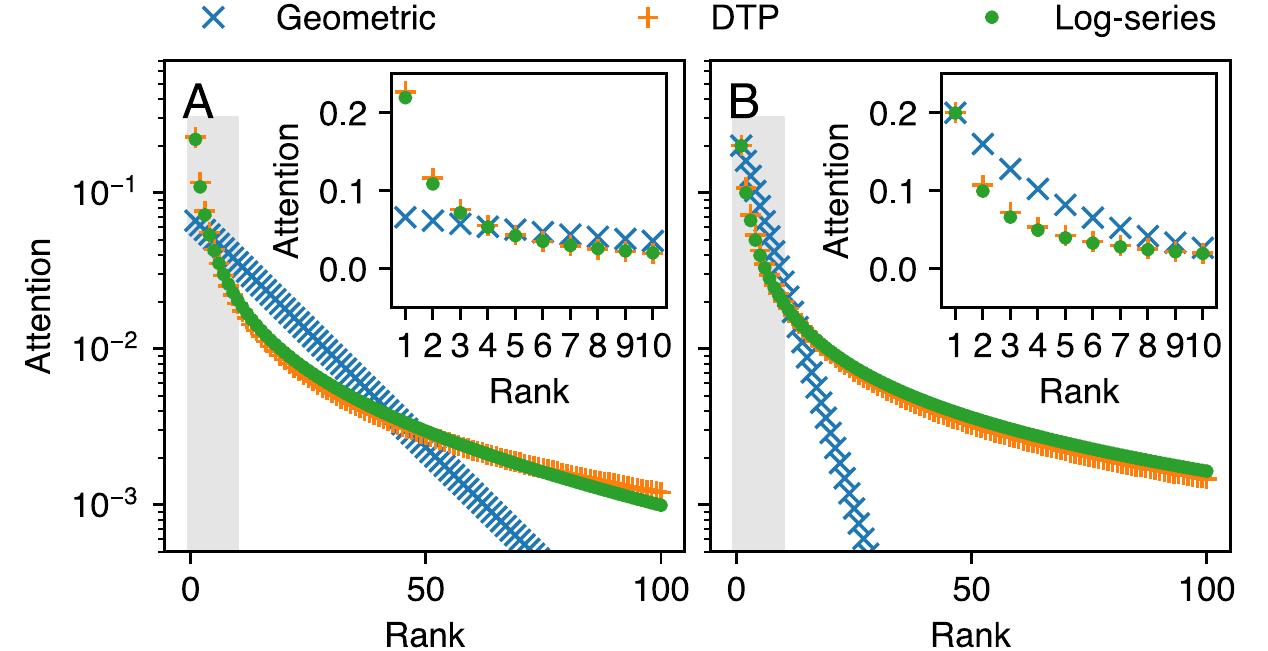}
     \caption{Comparison of Geometric, Log-series, and Discrete Truncated Pareto (TDP) on a list with length $n = 100$. In (A), the parameters are set so that $E[W_R]$ (\ie the mean number of results seen) is 15. In (B), the parameters are chosen so that $\omega(r_1) = 0.2$.} 
    \label{fig:distributions}
\end{figure}

Some families of distributions that fit the above criteria are presented in~\autoref{fig:distributions} and include Truncated Geometric Distribution, Truncated Log-series Distribution, and Truncated Discrete Pareto distribution~\cite{kozubowski2015discrete}. For the remainder of the paper we use the truncated geometric distribution but in an actual measurement scenario, another choice may be more appropriate.

\para{The Case of Small $n$.} The above distributions are applicable when $n$ is reasonably large, \ie well beyond the human attention span. When $n$ is small, we can draw inspiration from psychology: Miller's Law famously states that the average human's working memory can hold roughly 7 objects at a time~\cite{miller-1956-seven}. When $n$ is sufficiently small, we expect the user to read all of the results.\footnote{This may occur in practice when a user queries for obscure information or the vendor lacks data relevant to the search query} Formally, when $n$ is small, $W_R \dot\sim unif\{0, n\}$. Note that when choosing $W_R \defeq  unif\{0, n\}$ and $\hat{p} \defeq \bar{L_R}$, $E_R$ is precisely equal to $\hat{p}$; therefore all short lists exhibit approximate representational parity.

\para{The Problems of Inverse Log Scaling.} We notably left out inverse log:
\begin{equation}
    W_R = normalized\bigg(\bigg[\frac{1}{\ln(2)}, \dots, \frac{1}{\ln(n+1)}\bigg]^T\bigg).
\end{equation}
Inverse logarithmic scaling is commonly used in IR relevancy metrics such as nDCG, and also appears in some metrics of fairness in ranking~\cite{yang2017measuring}. However, this choice of $W_R$ has two major flaws: \textit{first}, it decays at a very slow rate and does not meet the relative convergence requirements described above in criteria (2). Even when $n = 1000$, $\frac{\omega(r_1)}{\omega(r_{1000})}$ is approximately $9.964$, which implies that the last 10 search results are, in aggregate, as influential as the first result. \textit{Second}, since there are no parameters for this choice of $W_R$, it incorrectly assumes that user behavior is the static across all platforms and for all search queries.\footnote{The base of the logarithm is irrelevant; after normalization, they all evaluate to the same vector of values.}

\subsection{Estimating Population Demographics \texorpdfstring{$\hat{p}$}{p hat}}
\label{sec:estimating-p-hat}

Our metric requires the specification of an estimator $\hat{p}$. This section describes how to choose $\hat{p}$ so that it serves as a reasonable prior for statistical parity.

\para{Implicit Estimators for $p$.} A tempting proposal would be to choose an \textit{implicit estimator} for $p$ based on intuition or observational data. 
This is plausible when the search query $Q$ is relatively simple. 
For example, consider the case when the user is querying a resume search engine for a list of certified nurses: recent data shows that the US national gender ratio for this profession is approximately 9.5:1 female to male~\cite{rappleye-2015-nurses}, thus we expect fair rankings for this query to reflect this. 
For more complex queries, however, demographic data may unavailable. If the user instead queries for Android Developers in Greenville, AL with at least three years of experience, we lack an external data reference. Thus, we are unable to choose nor justify an implicit estimator $\hat{p}$. 

Furthermore, $Q$ itself cannot be directly examined in many IR systems. Online services rank their feeds using proprietary algorithms that rely on personalization. From an auditor's standpoint, a generalized fairness metric must have the ability to estimate $p$ regardless of whether or not we have access to $Q$.

\para{Choosing $\hat{p} \defeq \bar{L_R}$.} An IR-motivated alternative to circumvent these issues is to determine $\hat{p}$ based on the vendor's data $R$. Suppose that within an IR system, the vendor evaluates $Q$ then filters their corpus to yield a subset of results $R$ in which all $r_i \in R$ meet some relevancy threshold. Then we can calculate $\hat{p}$ as defined in Equation~\ref{eq:phat}, \ie an equally-weighted sum $\forall l(r_i) \in L_R$. 
Since this calculation relies on the vendor's data $R$, the validity of $\hat{p}$ is dependent on the integrity of the vendor's data. Thus, it is imperative to first audit the vendor's \textit{data curation} for sampling bias and \textit{result scoring} for direct discrimination. A hypothetical example of unfair data curation would be a job site that refuses to add female software engineering candidates to their database. Unfair query evaluation could occur when a vendor has female candidates in their database but fails to show them when a recruiter queries for ``software engineers'' (\ie being a woman directly impacts the relevancy score). Fortunately, this style of audit is often straightforward: Chen~\etal tested for direct discrimination in scoring by posting two identical resumes that varied only by gender, and showing that they appeared at directly adjacent positions in search results~\cite{chen2018investigating}. If a preliminary audit finds that any of these assumptions are jeopardized, we can immediately deem $R$ as unfair without needing to calculate $\hat{p}$ or other components of Viable-$\Lambda$.

\para{Consequences.} In choosing $\hat{p} \defeq \bar{L_R}$, our prior for statistical parity is fitted to the vendor's data. Thus, if the vendor's knowledge of $p$ is lacking or biased (for example, unknowingly exhibits sampling bias), our estimate of $p$ will be as well. 
In this situation, Viable-$\Lambda$ becomes a metric of how well the vendor's ranking \textit{represents their own knowledge of $p$}.
% Another effect that follows, is that for a flat attention distribution function, all rankings will be deemed group-fair. 
%% ^ We can remove this line because we cover it in `The Case of Small n`

%%%%%%%%%%%%%%%%%%%%%%%%%%%%%%%%%
%%%%%   d and \delta_{min}  %%%%%
%%%%%%%%%%%%%%%%%%%%%%%%%%%%%%%%%

\subsection{Distance Metric \texorpdfstring{$d$}{d} and the \texorpdfstring{$\delta_{\textrm{max}}$}{delta min} Threshold}
\label{sec:d-and-delta-min}

We use $d$ and $\delta_{\textrm{max}}$ to delineate an \textit{acceptance region} around $\hat{p}$.

\para{Distance Metric $d$.} $d$ is a statistical distance metric that quantifies the difference between the two probability distributions $E_R$ and $\hat{p}$. %The following choices for $d$ would be appropriate in practical applications: (1) Total Variation Distance (TVD), (2) $\chi^2$ Independence Test, and (3) Numerical Methods.
The choice of $d$ follows naturally based on the domain of $E_R$ and $\hat{p}$. In Section~\autoref{sec:case_studies}, we demonstrate the use of Z-approximation for the binomial test statistic when $E_R$ and $\hat{p}$ are both binomial distributions.

\begin{figure}
    \centering
    \includegraphics[width=1\linewidth]{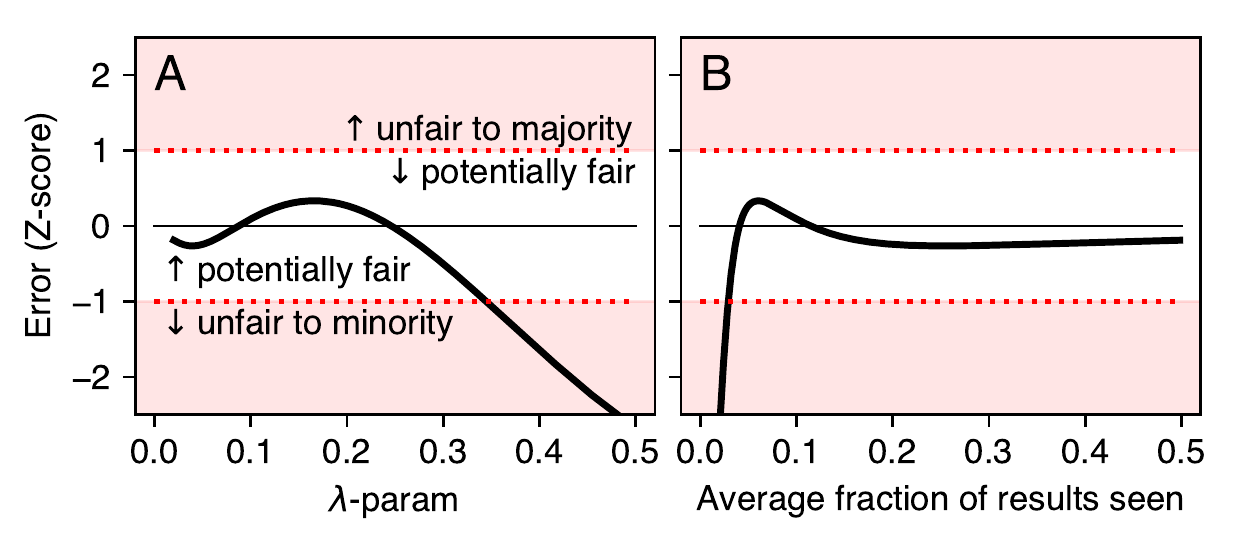}
    %\caption{}
    \caption{We find the range of the $\lambda$-parameter (and, thus, the corresponding average fraction of seen results (B)) for which the result list is potentially fair. If the attention given to a group is more than $1 Z$ away from its representation in the population, we deem the results unfair.}
    \label{fig:fitting_p}
\end{figure}

\para{Maximum Acceptable Distance $\delta_{\textrm{max}}$.} $\delta_{\textrm{max}}$ establishes an acceptable range of values around $\hat{p}$ in which we can safely assume that group fairness is preserved. In general, $\delta_{\textrm{max}}$ is the \textit{test statistic} threshold to the statistical significance test $d$. In conjunction, $d$ and $\delta_{min}$ constitute a statistical significance test with $h_0$: $E_R \sim \hat{p}$, \ie $E_R$ has the same sampling distribution as $\hat{p}$. A well-constructed $\delta_{\textrm{max}}$ takes the following factors into consideration:

\topic (1) \textit{Statistically significant difference between $E_R$ and $\hat{p}$ (size of $R$).} 
Suppose that we perform a search to yield results $R$ with length $n = 100$, for some binomial alignment property. % Our target property $p$ is a binomial distribution between classes $S$ and $S^C$. 
One interpretation of this scenario is that our result $R$ is a sampling distribution of size $n = 100$ from a universe $U$ of \textit{all} relevant candidates. We can formalize this by saying that $R$ is a simple random sample of $U$, where $n = 100 \ll |U|$.\footnote{Frequentist statistics relies on the assumption that $n \ll |U|$, which reflects the fact that it is often impractical or expensive to collect a census over the entirety of $U$.} $U$ has a true binomial parameter $p$. Thus, we can employ statistical inference to construct a confidence interval for $p$. For a binomial distribution, the maximum likelihood estimate (MLE) of $p$ yields $\hat{p} = \bar{L_R}$ with standard error $s_{\hat{p}} = \sqrt{\frac{\hat{p}(1-\hat{p})}{100}}$. By specifying a confidence level (\eg 95\% corresponding to $z=1.96$), we can build a confidence interval for $p$. Thus, we cannot reject $H_0$ if $E_R$ falls inside this range of values.
For categorical $p$, we can check for statistically significant difference between $\hat{p}$ and $E_R$ using the appropriate test statistics (\eg Pearson's $\chi^2$ Test for Independence).

\topic (2) \textit{Number of Search Queries Made.}  
This factor is applicable only in the case where the the vendor randomizes their search results. We can model $l(r_i)$ as the sampling distribution of $S$ across $k$ search realizations. We calculate the standard error in the same manner as above using $k$ in lieu of $n$.
An important distinction is that the previous standard error calculated with $n$ represents uncertainty in $p$, while this standard error calculated with $k$ represents uncertainty at each rank in $l(r_i)$. In the case where the vendor does not randomize their rankings or when the auditor can make an unlimited number of queries, this factor is irrelevant.\footnote{With an unlimited number of queries, we can arbitrarily reduce the standard error to any  $\epsilon > 0$. This is not possible with the confidence interval described in the previous section since it is the vendor---not the auditor---who determines the size of $R$. In the case where the vendor is able to increase $n$ (\ie by acquiring more data), then both the auditor and the vendor will be able to make more accurate estimates of $p$.}
    
% \topic (3) \textit{The Group/Individual Fairness Tradeoff.} Beyond the previous two terms, we can arbitrarily increase $\delta_{\textrm{max}}$ to allow leniency to the vendor. This can be a potential way to quantify the group/individual fairness tradeoff. However, since this paper focuses on group fairness, we will leave this as a potential avenue for future researchers.

\para{Plotting $d(E_R, \hat{p})$.} In the case where $W_R$ has one parameter, we can plot $d(E_R, \hat{p})$ as a function of $\lambda$. An example of this is illustrated in Figure~\ref{fig:fitting_p}.

\subsection{Restricting the Domain of the Parameter Space \texorpdfstring{$D$}{D}}
\label{sec:restrict-param-space}

In our model, we accept the null hypothesis that a ranking $R$ is fair if there exists a set of parameters $\Lambda$ within the parameter space $D$ which brings us within an acceptable range of parity. However, we are susceptible to type-II errors since not all $\Lambda \in D$ match realistic expectations about user behavior. Thus, by truncating our parameter space $D$, we can increase the \textit{power} (the probability of \textit{not} making a type-II error) of our hypothesis test. 

\para{Context-based Assumptions.} To illustrate this, consider a ranked list with length $n = 100$. Suppose that we have deduced that $W_R \sim Geom(\lambda, 100)$. The mathematically-permitted domain of the success parameter is $\lambda \in (0, 1)$. However, if we know for certain that  $E[W_R]$ is between 2 and 50 (\ie the average user views between 2 and 50 profiles), thus we can restrict our domain space to $\lambda \in (0.02, 0.5)$.\footnote{The expected value of $Geom(\lambda)$ is $1/\lambda$. Since $W_R$ is truncated, $E[W_R]$ asymptotically approaches this value for large $n$.}

\para{Empirically-informed Truncations.} Furthermore, we can also set up a small-scale experiment to estimate $E[W_R]$. Suppose that we have a small group of users with $N=16$. Assume that the average number of results viewed is approximately normal with $\bar{x} = 27$ and standard error $s = 12$. Using the maximum likelihood estimator (MLE) we can construct a 95\% confidence interval for $\bar{x}$ between approximately 21 and 33. Since $E[W_R] \approx 1/\lambda$, we can say that the parameter $\lambda$ lies between 1/21 and 1/33, corresponding to the interval $(0.03, 0.47)$\footnote{We can use the confidence interval of $\mu$ to restrict the parameter space of other families of probability distributions by replacing $1/\lambda$ with the corresponding equation for $E(W_R)$.}.

As we can see in the above example, it is possible to leverage data from small-scale experiments to estimate the likelihood of $\lambda$. This offers us an avenue to improve the statistical power of the Viable-$\Lambda$ test.\footnote{This approach is different from the eye-tracking studies mentioned in~\autoref{sec:W-R}. Instead of using eye-tracking heatmaps to construct the entirety of $W_R$, use a singular dimension of the data (namely $E[W_R]$) to inform a reasonable range of values for $\Lambda$.}

\subsection{Generating Fair Ranked Lists}

\begin{figure}
    \centering
    \includegraphics[width=1\linewidth]{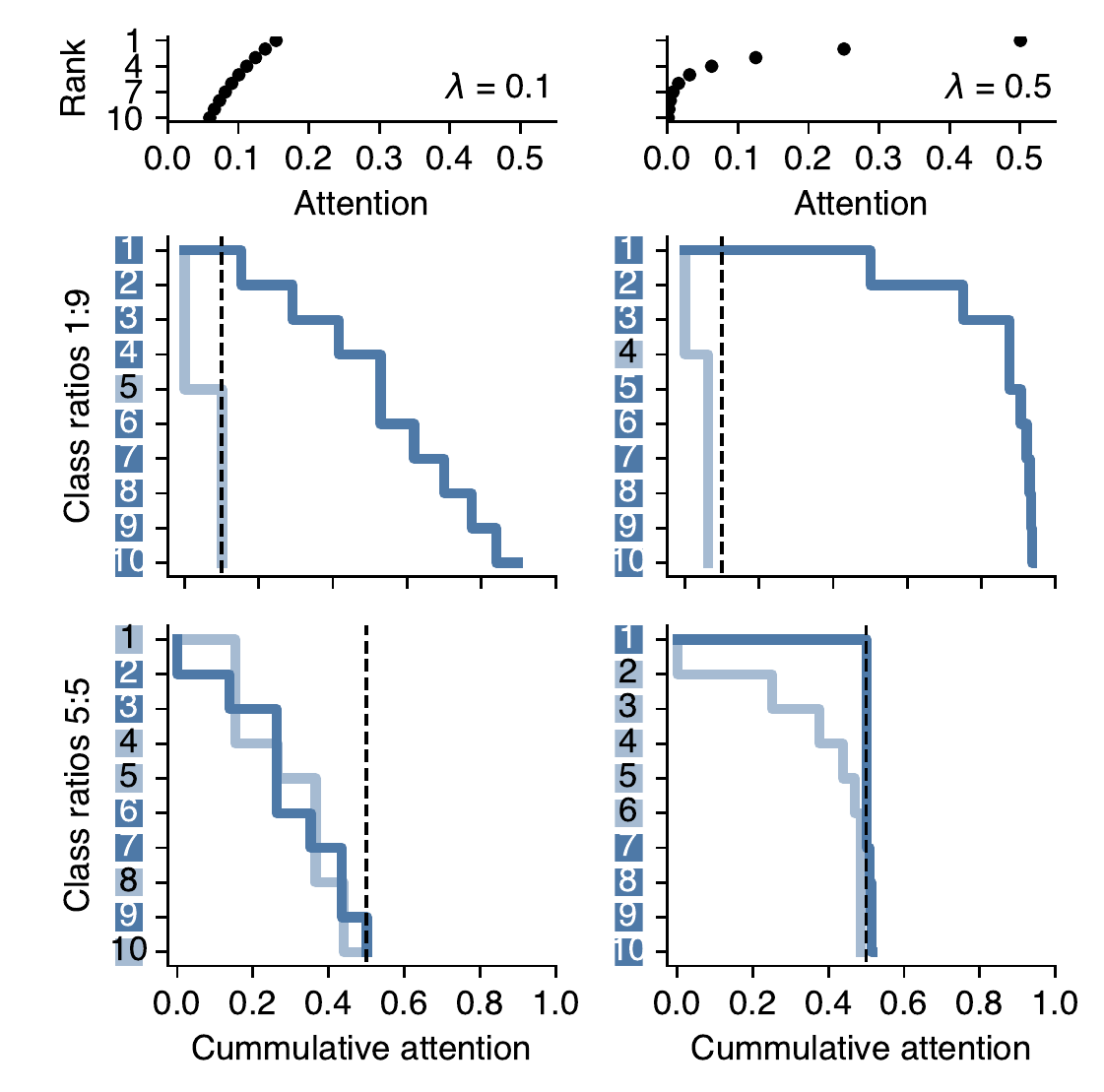}
    
    \caption{We generate rankings that satisfy our definition of attention fairness. Note that depending on the attention distribution function, fair rankings with the same proportions of classes differ.}
    \label{fig:synthetic_fair}
\end{figure}

Other researchers have focused on creating fair ranked lists and their cost in terms of individual fairness in depth~\cite{zehlike2017fa,celis-2017-arxiv,singh2018fairness,biega2018equity}.
Here, we provide a few examples of ranked lists that are fair for fixed $\lambda$. 
% For a clear illustration, we assume all items are equally relevant and we present rankings that attain representational parity.
Figure~\ref{fig:synthetic_fair} shows four best-attempt fair lists with varying class imbalance (1:10 in top row, 5:10 in bottom row; minority class A with light blue, majority class B with dark blue) and attention distributions ($\lambda=0.1, 0.5$ from left column to right).
In the top row, we see that as the distribution becomes steeper, the one minority sample is placed higher to receive proportional attention.
In the bottom row, the flattest distribution requires a list of results were both classes are quite mixed.
However, in the case of the steepest distribution, all elements of class B are placed in a block from rank two on to match the attention already given to class A by its representative at first rank.

\section{Case studies}
\label{sec:case_studies}
%To demonstrate the properties of the Viable-$\Lambda$ Test, we apply it in {\em simulated audits} of three ranking services: a hiring site used by recruiters to search for employees, a dating service, and a web search engine.
%Note that we only use them here to demonstrate how the metric could be applied to real-world services.
%Thus, we do not claim that the presented results are definitive descriptors of fairness of algorithms behind these services.

In this section, we apply the Viable-$\Lambda$ Test to three different search engines: (1) gender fairness on a hiring site used by recruiters to search for candidates, (2) racial fairness on a dating site, and (3) political fairness on a search engine.
As a disclaimer, we are only using these results to demonstrate a use of our metric on real-world data. We do not \textit{not} claim that any of these are biased or unfair.

\para{Ethics of Data Collection. } 
While conducting our measurements we were considerate both of the services we collected the data from and of people who this data represents. 
All collected data is available to any person with an account on the corresponding service.
We did not interact with any users of these services as part of the collection. 
Additionally, we minimized any impact on the operations of the services by using a low query intensity (at most one query every 30 seconds).  Finally, we adhered to the usage quotas of Face++.\footnote{\url{https://www.faceplusplus.com}}

\para{Assumptions}
In our case studies, we assume that $W_R \sim Geom(\lambda)$. As discussed in Section~\autoref{sec:W-R}, $Geom(\lambda)$ meets the desired properties of an attention distribution, and thus serves as a reasonable starting approximate for the true $W_R$.

We used \href{http://www.statistics4u.com/fundstat_eng/cc_optim_meth_brutefrc.html}{brute-force optimization} over our one-dimensional parameter space $D$ to check if $\exists \lambda \in D$ such that $d(E_R,\hat{p}) < \delta_{\textrm{max}}$.

% TODO: Mention Binomial p, d, and delta_max here once we switch to 2*Z

\subsection{Gender in Hiring} 
%We begin by examining ranking of worker profiles in a resume search engine.  The dataset used in this case study was collected and made available to us by Chen~\etal~\cite{chen2018investigating}. Using the recruiter's interface to the service, the Chen~\etal queried 35 different search terms (such as ``bartender'', ``electrical engineer'', ``laborer'', ``pharmacist'', or ``software engineer'') in 20 US cities, resulting in 692 non-empty result lists, including 412 with length of least 100 results. Then, they inferred the gender of each candidate from the given name (see \cite{chen2018investigating} for details).
%The gender estimate is a real number describing the probability of a candidate being male.
%In the original work the data was dichotomized, assuming `male' for probability $>$0.8 and `female' for probabilities $<$0.2; remaining 8\% of profiles were omitted.

Our first cast study examines the ranking of job candidates on a resume search engine. Data was collected and made available to us by Chen~\etal~\cite{chen2018investigating}. Using the recruiter's interface to the service, the authors queried 35 different search terms (such as ``bartender'', ``electrical engineer'', ``laborer'', ``pharmacist'', and ``software engineer'') in 20 US cities, resulting in 692 non-empty result lists. Of these results, we look at the 412 with length $n \geq 100$. 

\para{Modeling Gender Alignment $L_R$}
Chen~\etal determined the gender of each candidate based on their given name (see \cite{chen2018investigating} for details).
The gender estimation is a real number describing the probability of a candidate being male between 0 (female) and 1 (male).
The original authors dichotomized the data, assuming `male' for probabilities $>$0.8, `female' for probabilities $<$0.2, and omitted other profiles. These constituted 8\% of all candidates.
In this study, we use the gender alignment probabilities directly. We do neither omit ambiguous candidates, nor project alignments to their most-likely class.

\para{Determining $\hat{p}$}
In this study, the true population demographics for each query are unknown. As per Section~\autoref{sec:estimating-p-hat}, we cannot choose an \textit{implicit estimator} and instead leverage the vendor's data by choosing $\hat{p} \defeq \bar{L_R}$ in each query.

\para{Evaluating $d$ and $\delta_{max}$. }
$\hat{p}$ and $E_R$ are both binomial distributions representing gender alignment. To check for statistically significant difference between them, we use the Z-test approximation for the binomial test. This $d(E_R, \hat{p}$ is our test statistic. We deem $R_i$ unfair if $\nexists \lambda$ such that this test statistic is less than 1Z. This represents a ~68\% confidence that fair representation is impossible.

\begin{figure}[t!]
    \centering
    \includegraphics[width=1\linewidth]{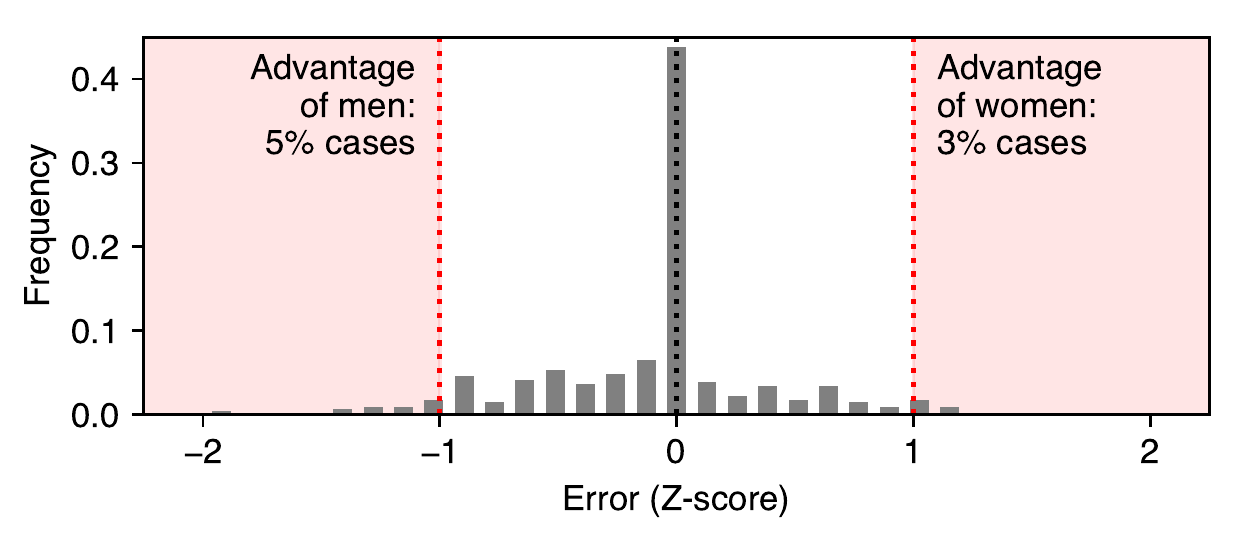}
    \caption{Best case scenario of error distribution among location-job pairs in the resume search engine. For each location-job query we attempt to find a $\Lambda$ for which the ranking could be considered fair. We fail to do so in 8\% of cases.}
    \label{fig:fitting_e}
\end{figure}

\para{Viable-$\Lambda$ in Hiring. }
In Figure~\ref{fig:fitting_e}, we run Viable-$\Lambda$ for each list and plot the minimum attainable $d(E_R, \hat{p}$ for each $R_i$. About 92\% of rankings can be considered fair at the 1Z threshold. Still, 6\% of these lists appear biased against women and 3\% appear unfair to men regardless of the distribution function.
All rankings are deemed fair at the 2Z threshold

% \textit{First}, we try to find a $\lambda$-parameter of the geometric distribution of attention independently for each location-job pair to see if they could be considered group-fair (without considering relevance).
% We derive the population estimator separately for each ranking, along with its standard error and maximum allowable distance of $\delta_{\textrm{max}}=1Z$.
%As presented in Figure~\ref{fig:fitting_e}, the vast majority of result lists could be considered fair, \ie there exist such an attention function in which the attention given to each gender is within $1Z$ of $L_R$. 

\begin{figure}[t!]
    \centering
    \includegraphics[width=1\linewidth]{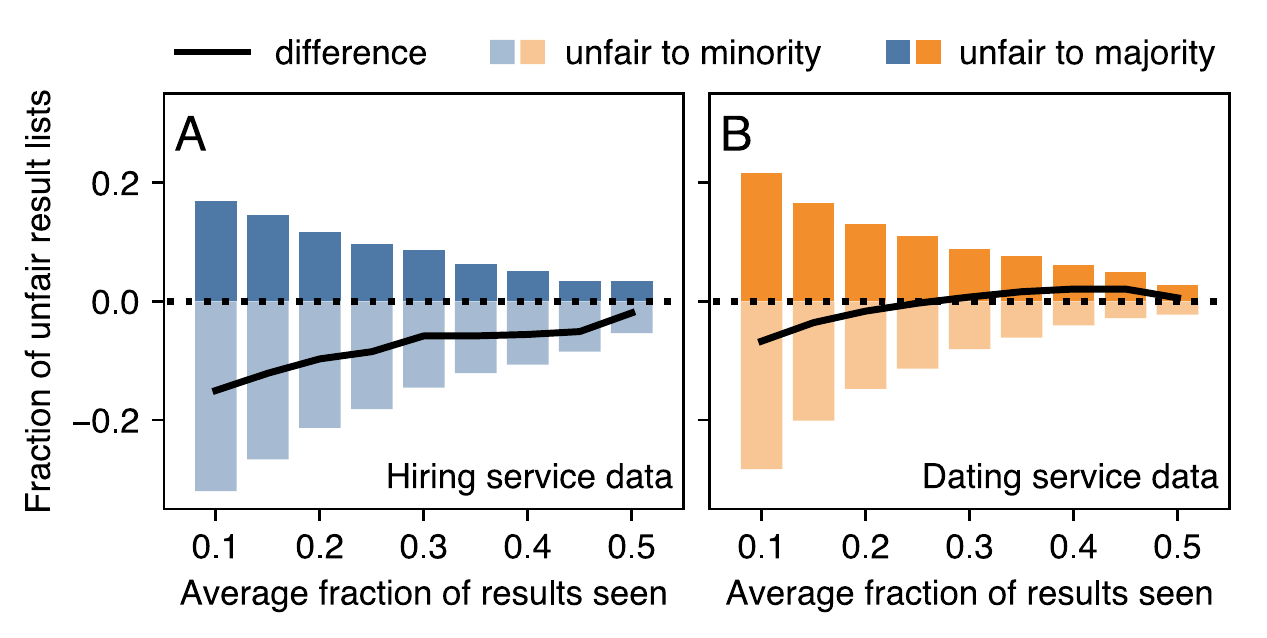}
    \caption{The more results a user sees, the higher the chance that the ranking can be considered fair (in extreme cases when all results are seen, the ranking is always fair). A) Regardless of $\lambda$, women are underrepresented more often than men in the job-location searches. B) Depending on the true value of $\lambda$, Black profiles may be underrepresented more frequently than non-Black profiles. }
    %\caption{Depending on the usage of the search engine under consideration, the auditor controls the mean number of results seen. The more results a user sees, the higher the chance that the ranking can be considered fair (in extreme cases when all results are seen the ranking is always fair). A) Regardless of attention parameterization, women are unfairly presented in a greater fraction of job-location pairs than men in the hiring dataset. B) Depending on the usage patterns in the dating service, profiles of Black users could be getting more or less attention than those of non-Black users.}
    \label{fig:hiring_dating}
\end{figure}

In Figure~\ref{fig:hiring_dating}A, we demonstrate the effect of $\lambda$ on $d(E_R, \hat{p})$.
As per Section~\autoref{sec:restrict-param-space}, we assume within reason that $E[W_R] \in [.1n, .5n]$ in the context of this search engine.
The horizontal axis depicts nine $\lambda$-parameter choices corresponding to a user viewing between 10\% and 50\% of all results on average. 
For all sampled values of $\lambda$, rankings under-represent women more often than men.

%\textit{Second}, rather than fitting a separate attention distribution function to each result list, we test all lists against nine different $p$-parameter values that correspond to the users seeing on average between 10\% and 50\% of the ranks.
%The results are shown in Figure~\ref{fig:hiring_dating}A. 
%We note that across all values of the $p$-parameter, there is a larger fraction of result lists that give less attention to female profiles than one would expect from the fraction of the workforce they constitute.

While the gender ratio is balanced in the dataset, the gender proportions vary widely between queries. We calculated the $\hat{p} = \bar{L}$ estimate for $p$ for each query separately. Although the vast majority of rankings passed the Viable-$\Lambda$ test, rankings tended to under-represent women more frequently for most values of $\lambda$.

\subsection{Race in Dating}
% We used one of the authors personal account to obtain lists of dating profiles by querying an online dating service's API in the same way the browser would, once every 15 minutes for a week. 672 sessions of 100 results each were recorded, containing 4,407 unique profiles. Even though we were running the same query repeatedly, we observed significant churn in the results, with a fraction of profiles being replaced and reordered every time. We used Face++ to infer the race of each person from their profile picture. Because the data is only used for illustration purposes, the precision of the race detection is not crucial in these examples.\footnote{Face++ inferred a different gender than self-reported for 13\% profiles identified as white, 11\% profiles identified as Asian, and 8\% of profiles identified as Black (previous studies showed disproportionately high gender misclassification rates for Black women~\cite{buolamwini2018gender}). }

We used an author's personal account to query an online dating service's API. The script ran once every 15 minutes over the course of one week, collecting 672 lists each with length $n = 100$. We determined that each of these lists sampled from a pool 4,407 unique profiles. Even though we were running the same query repeatedly, we observed significant \textit{shuffling} and \textit{churn} in the results, with some profiles being replaced more often than others.

\para{Modelling Race Alignment. }
We used Face++ to infer the race of each person from their profile picture. Because the data is only used for illustration purposes, the precision of the race detection is not crucial.\footnote{Face++ inferred a different gender than self-reported for 13\% profiles identified as white, 11\% profiles identified as Asian, and 8\% of profiles identified as Black. Previous studies have shown high gender misclassification rates for Black women in particular~\cite{buolamwini2018gender}). }
%In 813 (18.4\%) profiles, Face++ did not detect exactly one face; among the identified profiles 2,411 (54.7\%) were classified as white, 573 (13.0\%) as Black, 534 (12.1\%) as Asian, and 76 (1.7\%) as Indian. 
%The classifier we used only returns one race label per query and does not report its certainty, thus we treat this assignment as binary: Black or non-Black.
In 813 (18.4\%) profiles, Face++ detected $>1$ face; among the identified profiles 2,411 (54.7\%) were classified as white, 573 (13.0\%) as Black, 534 (12.1\%) as Asian, and 76 (1.7\%) as Indian. 
For each profile photo, Face++ returns only the \textit{most likely} race, not its certainty. 
One potential option is to model alignment as a categorical probability distribution among the five aforementioned classes. In doing so, we would use the $\chi^2$ Independence Test to compute the statistical difference between $E_R$ and $\hat{p}$. This would allow us measure the representation of all classes simultaneously. To simplify our illustrations, we project our alignment into a binomial distribution - \ie Black vs. non-Black profiles.

\para{Determining $\hat{p}$, $d$, and $\delta_{max}$. }
In this example, we ran the same query across each of our searches, yielding 4,407 unique results that meet the vendor's relevancy threshold. In our first approach, we chose $\hat{p} \defeq \bar{L}$ on the set of all unique results. In doing so, we can build a 95\% confidence interval for $p$; namely $\hat{p}$ is a \textit{binomial sampling distribution} with $n = 4407$. Since we have a large sample size, our test statistic has a very small error threshold. We use the Z-test approximate for binomial test as our distance metric $d$ again. Using this definition of $\hat{p}$, Viable-$\Lambda$ showed that rankings unequivocally over-represented Black profiles. While Black profiles constituted only ~13\% of all unique profiles, the average ranking displayed ~16\% Black profiles. 

\para{Potential for Correcting Societal Bias. }
%There are
%Why might a dating service want to manipulate the representation of certain groups? 
% Up to this point we used the average fraction of Black users in the lists as the population estimator $\hat{p}$.
% However, while Black users' profiles constitute $\sim$13\% of all unique profiles, they correspond, on average, to $\sim$16\% of profiles shown in the ranks. 
% Because we only collected the data from the point of view of one user in one location, we are unable to say whether this is due to differences in frequency of use of the service, personalization, location effects, or a conscious decision by the service provider.
Black users have been shown to be disadvantaged in online dating~\cite{rudder2014dataclysm}.
Thus, it is possible that these profiles are over-sampled by this dating service to compensate for their lower click-through rates. 
% Our metric can use an arbitrary $\hat{p}$ estimator to account for correction and compare the attention to the given value, rather than the true underlying population.
Suppose that we want to measure how well rankings match the vendors' \textit{manipulated} population demographics.  
Then in this case, $p$ is the true percentage of Black profiles displayed by the vendor's ranking algorithm. At each rank $k$, we have a binomial sampling distribution of Black profiles with size $n = 672$ (i.e. the total number of searches). Thus, we can continue using the Z-test approximation for binomials as our distance metric.
%Since $n \gg m$, $n$ has negligible impact on $\delta_{max}$, and we can use ;\textit{}% based on the number of samples, we can assume $\hat{p} \approx p$. 

\para{Evaluating $R_i$ Individually. }
We begin by evaluating each ranked list $R_i$ individually.
%Given the context of our search engine (as per \autoref{sec:restrict-param-space}), we can reasonably assume that $E[W_R] \in [10, 50]$. 
%In Figure~\ref{fig:hiring_dating}B, we illustrate the effect of $\lambda$ on $d(E_R, \hat{p})$. The figure depicts nine $\lambda$-parameter values that correspond to the users seeing on average between 10\% and 50\% of all results. 
First, we find that for all values of the $\lambda$-parameter, the majority of lists can be considered fair, see Figure~\ref{fig:hiring_dating}B.
For the steepest distributions (smallest value of the $\lambda$-parameter) more result lists are unfair towards the Black users than towards non-Black users;
when we assume an attention distribution function such that the users see on average 30\% to 45\% of results, there are more realizations in which Black users get more attention than proportional to their representation, and the situation equalizes for the least steep distributions.
This effect is caused by the fact that white profiles appear on the first position in 54\% of realizations, even though they only constitute 48\% of the observed population on average, but for the rest of the high ranks, Black profiles are presented more often than the population estimator would indicate.

\begin{figure}[t!]
    \centering
    \includegraphics[width=1\linewidth]{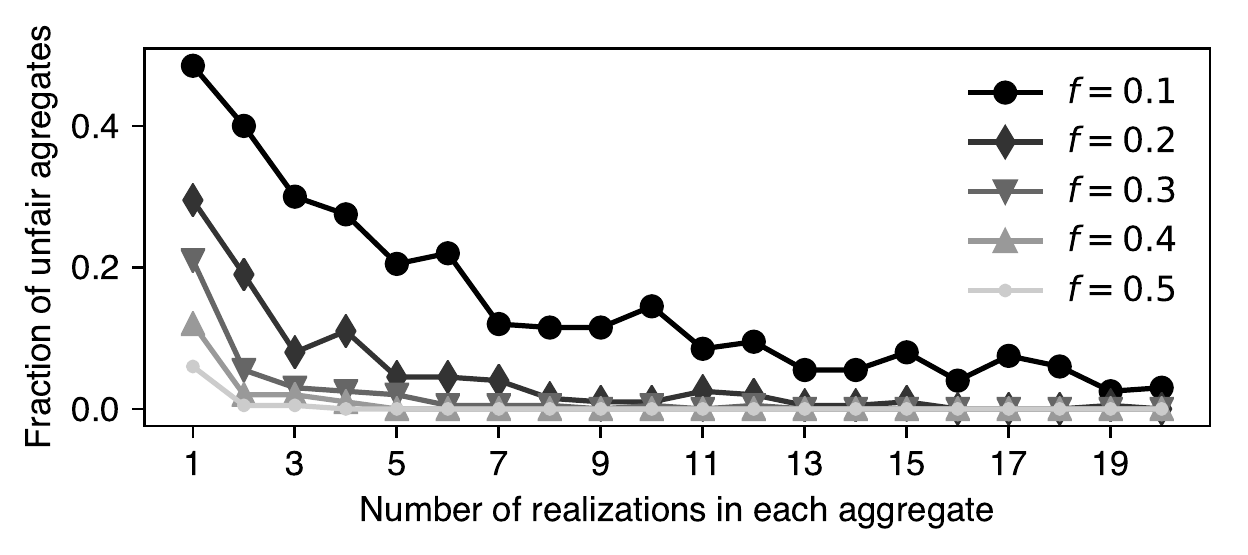}
    \caption{The investigated dating service returns a different set of results to the same query issued by the same person. As a consequence, while each ranking might be unfair, the results in aggregate are fair. The number of runs necessary to expect a fair ranking depends on the chosen distribution parameters.}
    \label{fig:dating}
\end{figure}

\para{Evaluating $R_i$ in Aggregate. } 
Next, we evaluate the fairness of several realizations sampled in aggregate. A similar notion was proposed by Biega~\etal~\cite{biega2018equity}.
As shown in Figure~\ref{fig:dating}, even if each single ranking realization is unfair, the aggregate of multiple unfair rankings can be considered fair.
Our metric can capture this because the alignment vector is not binary.
We note that the more realizations included in the aggregate, the higher the fraction of fair aggregates, regardless of the assumed attention distribution function.
Still, the steeper the function, the more runs are necessary to achieve a fair aggregate.
In a hypothetical case of multiple rankings generated by a random (and, thus, unbiased~\cite{yang2017measuring,zehlike2017fa}) ranker, each rank will converge to contain a proportional representation of classes, and the ranking will be fair regardless of the assumed attention distribution function.

\para{Summary. } This case study highlights several important findings. The perception of the existence and even the direction of bias depends on the attention distribution. Furthermore, the bias can be corrected over time by reshuffling the results. Finally, our metric can accommodate population estimates that are different from the underlying populations for example to correct for societal biases. 

\begin{figure*}[htb!]
    \centering
    \includegraphics[width=1\linewidth]{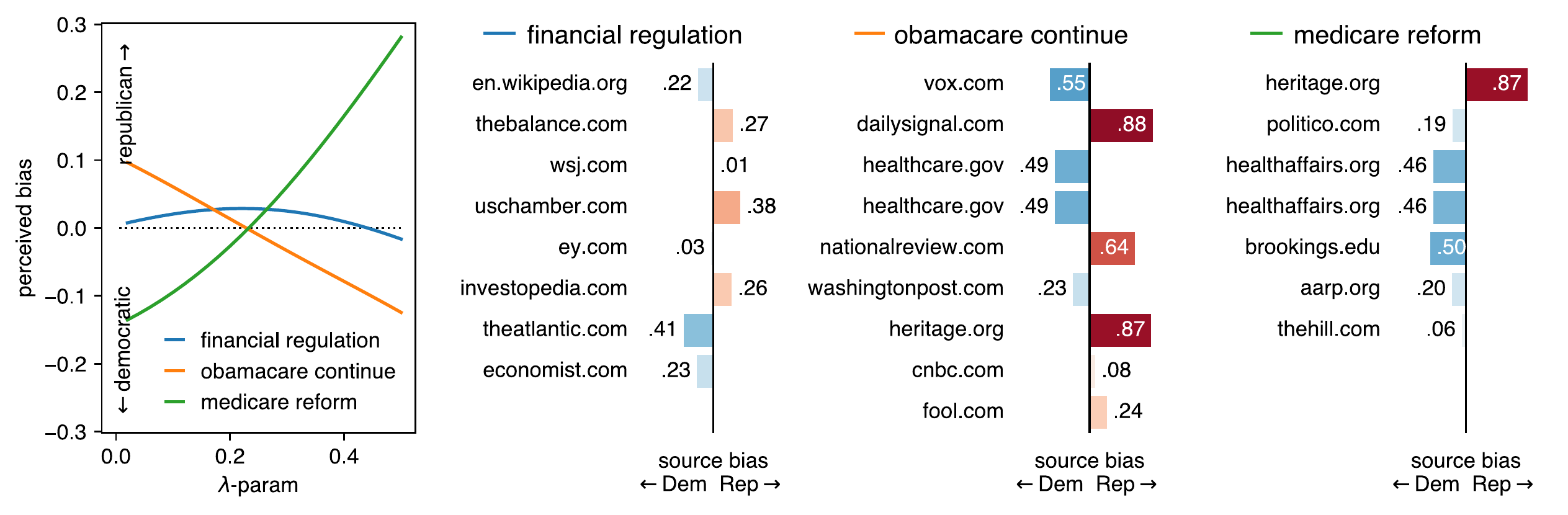}
    \caption{The perception of bias may depend on how the users of a service distribute their attention to the presented results. The neutral results for ``financial regulation'' form a list that appears neutral regardless of usage patterns. However, ``obamacare continue'' and ``medicare reform'' might appear partisan in either direction, depending on the attention distribution function.}
    \label{fig:partisan}
\end{figure*}

\subsection{Political Bias in Web Search}
Our third case study differs from the previous two in that there is no protected group; instead, the \textit{alignment} $l_i$ is a proxy for political leaning of each item in the ranked list.
The dataset we use was collected and made available to us by Robertson~\etal~\cite{robertson2018auditing}.
There are two elements to the dataset: (1) search results and (2) partisan audience bias estimates. 
The first part comprises the first pages of results to 1,443 different web search queries.
The second part maps 19,022 domains that appear in the search results to the bias scores on a liberal/conservative axis. 
Based on the tweets from registered voters, it assigns a number from -1 (only Democratic voters share content from that domain) through 0 (Democratic and Republican voters are equally likely to share content from this domain) to 1 (shared only by Republican voters).
For example, \texttt{blacklivesmatter.com} scores -0.94, \texttt{en.wikipedia.com} scores -0.22, \texttt{dhs.gov} scores -0.01, \texttt{youtube.com} scores 0.13, and \texttt{catholics4trump.com} scores 0.98.
Note that the score is assigned to a domain, not a specific webpage.

In this case study, we measure whether the aggregated partisan bias of search results is cancelled, given the attention distribution.
%Note that it is not feasible to unequivocally determine the existence of partisan bias in the results. Instead, 
Among the 1,443 searches in the dataset, we select three examples that best highlight the importance of considering attention distribution in the audit.
In the interest of brevity, we omit the distance metric steps of Viable-$\Lambda$. We do, however, report the difference between the source biases weighted by the attention per rank and 0. Positive values indicate that a result list leans conservative; negative values indicate liberal lean.

Figure~\ref{fig:partisan} presents the search results to three queries: ``financial regulation'', ``obamacare continue'', and ``medicare reform'', along with perceived partisan bias of each of these lists for different attention distribution functions.
Note that in the first panel, small values of the $\lambda$-param correspond to a flatter distribution; as the $\lambda$-param grows, more attention is given to top results.
The search engine might return multiple items from the same domain, therefore some domains appear multiple times in the list (for example \texttt{healthcare.gov} in Figure~\ref{fig:partisan}b).

The results in Figure~\ref{fig:partisan} for ``financial regulation'' are neutral regardless of how attention is distributed since most of the results are apolitical.
The results for ``obamacare continue'', on the other hand, lean republican: they feature three items from highly conservative sources.
Still, because the first result is liberal-leaning---and among the first four results, three lean liberal---the result list will appear to be liberal-leaning overall if the attention distribution function is steeper.
Finally, the results for ``medicare reform'' are almost exclusively from democratic-leaning sources. However, because the top result comes from a strongly conservative outlet, the list exhibits strong conservative-bias if the attention distribution function is steep. 
Thus, these examples illustrate how the shape of the attention distribution function can dramatically alter conclusions about the fairness/bias of ranked outputs.

\balance
\section{Limitations}
\label{sec:limitations}
\para{The Shape of $W_R$. }
While our framework allows for arbitrary families of $W_R$ , we only considered the truncated geometric distribution in our examples. The results of the Viable-$\Lambda$ test rely on an accurate model of human attention; thus further research into human perception and the quantification of the SEME would improve the basis of this metric.

% We also made it clear that without knowing these functions, we are unable to determine whether a ranking is fair---both a crucial finding, but also a limitation to our work.
% As operators implement fairness measures there may appear a need for more complex functions describing attention distribution than geometric or logarithmic.
Additionally, singlely-parametered $W_R$ may not be sufficient in modelling expected attention. One important parameter is pagination; researchers have found that the CTR of the last result on a page getting more clicks than the pre-to-last~\cite{murphy-2006-clicking}, and each page introducing a disproportional drop-off of attention~\cite{insights2013value}.
Furthermore, modern search engines often add variation into their search results.
For example, Google search might display the actual content the user is seeking directly in the result page, or group results by type (\ie ``Sponsored'', ``Video'', etc.).

\para{Population Estimators. }
In this work, we derived the population estimators $\hat{p}$ directly from $R$ using the $\bar{L}$ estimator.
Hence, we assumed that the items the vendor chooses to show in the top $k$ results are a proportional representation of all $N$ potential results.
It is likely, however, especially with large $N$, that they are not.
For example, a real-world candidate ranking system employed by Amazon was shown to systematically rank women lower than men~\cite{dastin-2018-amazon}.
If we audited it and only had access to the top 100 results out of 1000, we would be likely to assume a $\hat{p}$ that underestimates the fraction of female candidates.

\section{Discussion}
\label{sec:discussion}
Studies have shown that swapping search results can cause significant changes in users' eye-gaze and click patterns.
For example, after reshuffling Google Search results, unwitting users still tend to click on the top results, but some do shift more attention to lower ranks~\cite{keane2008people}.
While we can use our metric to check if $\Lambda$ exists such that representational parity is achieved (or---from the operator's point of view---verify that the results are fair given the known $\Lambda$), it is not guaranteed that altering results based on the metric's measurements will create fair rankings, since user behavior may change in uncertain ways. 
One possible way to create fair rankings is by means of a continuous, iterative process: the operator reshuffles the results to achieve parity under a measured $\Lambda$, users potentially change their behavior as a response, the operator updates the $\Lambda$ estimate, and so on.

\section{Conclusion}
\label{sec:conclusion}
In this work we introduced a novel metric of group fairness in ranked lists, tying the measurement to the consumers' attention distribution. 
We showed how our approach could be used by auditors on three real world examples. 
Our results highlight the need for modelling attention specifically for the audited service: depending on the attention distribution function, the same list of results can appear biased both in favor and against the protected group.
All code will be made publicly available upon publication.

\begin{acks}
  This research was supported in part by \grantsponsor{1}{NSF}{https://www.nsf.gov/} grant \grantnum{1}{IIS-1553088}. Any opinions, findings, and conclusions or recommendations expressed in this material are those of the authors and do not necessarily reflect the views of the NSF.
\end{acks} 

\bibliographystyle{ACM-Reference-Format}
\bibliography{references}

\end{document}